\documentclass[%
twocolumn,
 amsmath,amssymb,
 aps,
prb,
]{revtex4-1}

\usepackage[utf8]{inputenc}
\usepackage[T1]{fontenc} 
\usepackage[english]{babel}
\usepackage{graphicx}% Include figure files
\usepackage{dcolumn}% Align table columns on decimal point
\usepackage{bm}% bold math
\usepackage{amsmath}
\usepackage{amssymb}

\begin{document}

\title{Physical properties of MgO at deep planetary conditions}

\author{R. Musella$^1$, S. Mazevet$^{1,2}$, F. Guyot$^3$}
\affiliation{$^1$LUTH, Observatoire de Paris, PSL Research University, CNRS, Universite Paris Diderot.}%Lines break automatically
\affiliation{$^2$CEA-DAM-DIF,91280 Bruyeres le ch\^atel, France}
 \affiliation{$^3$Institut de Min\'eralogie de physique des Mat\'eriaux et de Cosmochimie (IMPMC), Museum National d'Histoire Naturelle, Sorbonne Universit\'e, IRD, CNRS, Paris, France}

\date{\today}

\begin{abstract}
  Using {\sl ab initio} molecular dynamics simulations, we calculate the physical properties of MgO at conditions
  extending from the ones encountered in the Earth mantle up to the ones anticipated in giant planet interiors such as Jupiter. We
  pay particular attention to the high pressure melting temperature throughout this large density range as this is a key ingredient
  for building accurate planetary interior models with a realistic description of the inner core.  We compare our simulation
  results with previous {\sl ab initio}  calculations that have been so far limited to the pressure range corresponding to the Earth mantle and the
  stability of B1-B2 transition around 6 Mbar. We provide our results for both the EOS and high pressure melting curve in parametric forms
  for direct use in planetary models. Finally, we compare our predictions of the high pressure melting temperature with various interior profiles to
  deduce the state of differentiated layer within the core made of MgO in various types of planets and exoplanets.
  
\end{abstract}
\maketitle

\section{Introduction}
As a major constituent of the Earth Mantle, the physical properties of magnesium oxide up to about 1Mbar (1Mbar=100GPa) have been
the subject of extensive studies on both the experimental and theoretical sides\cite{duffy}. For decades, this was especially the case for the
equation of state (EOS) and the high-pressure melting temperature that are key ingredients to model the composition of the (Mg,Fe)O ferropericlase component
in the Earth and allows us to put constrains on the solid or liquid nature of the lower mantle\cite{2005PhRvL..94w5701A}. In this pressure range relevant
to the physics of the Earth mantle, MgO is found to be remarkably stable in the B1 (NaCl) structure (periclase) with some discrepancies noted between
the calculated and measured melting temperature\cite{2010PhRvB..81e4110B}.

With the continuous discovery of terrestrial exoplanets several times the size of the Earth there is now a pressing need to extend
our knowledge of the phase diagram of a few key planetary constituents up to several Mbar and well beyond conditions encountered in the Earth mantle\cite{duffy}.
In particular, the modeling of super-Earths, whose density is similar to that of the Earth but with mass up to ten times the one of the Earth requires, for example,
to extend our knowledge of the phase diagram of MgO up to 15 Mbar\cite{2007ApJ...665.1413V,2007ApJ...656..545V,2008PhST..130a4035S}. This jump by an order of magnitude
of the pressure range that needs to be described is a significant challenge for both experiments and theoretical methods. 

In addition to the large amount of experimental data and theoretical work reported for periclase at mantle conditions\cite{duffy}, laser-shock compression has
been recently used to constrain the mgO phase diagram well beyond Earth mantle conditions. A first set of experiment identified discontinuities in the shock
velocity data as a transformation from the B1 (NaCl) to the B2 (CsCl) structure at 4.5Mbar and 5,000K and 7,000K and melting at 6Mbar and about 12,000K\cite{2012Sci...338.1330M}
using reflectivity measurements and assuming metalization upon melting. This B1-B2 transition was previously predicted by density functional theory (DFT) calculations
and corresponds to an increase in coordination number of the Mg atoms from 6 to 8 atoms of oxygen\cite{2005PhRvL..94w5701A}. These indirect constrains on the
high-pressure phase diagram were followed up by high-power laser experiments that identified the B1-B2 transition directly using diffraction measurements and
ramp loading to remain well below the temperature reached upon direct shock while reaching similar pressures\cite{2013NatGe...6..926C}. These series of experimental
measurements provide a first sketch of the high-pressure phase diagram of mgO with a B1-B2 transition occurring between 4 and 6 Mbar with a negative Clapeyron slope,
a B2 structure stable up to 9 Mbar and melting anticipated to occur at around 12,000K at 6Mbar. This pioneer work has triggered intense experimental and theoretical
activities for the past few years and there are currently heated debates regarding the exact pressure-temperature location of these phase transitions up to below
7Mbar\cite{2014PhRvB..89m4107C,2015PhRvL.115s8501R,2016GeoRL..43.9475B,2015PhRvE..92b3103M,2013PhRvL.110m5504B}.

In the present paper, we use DFT simulations to significantly extend the phase diagram of MgO to encompass the conditions anticipated not only in Super-Earths
of up to 10 Earth mass (~15Mbar) but also to the ones expected in the cores of icy giant and giant planets (~100Mbar). MgO is indeed one end-member component 
in the core of these more massive planets and could even be directly present as one dissociation product of the stable silicates MgSiO$_3$ or Mg$_2$SiO$_4$ that
constitute planetary embryos in the core accretion model\cite{1996Icar..124...62P}. We paid particular attention to the high-pressure melting curve calculated across
this up to now unexplored pressure range as this is required for the modeling of giant planets by providing an additional constrain on the nature of the inner core.
We also revisit the pressure region around 5Mbar and compare our results for the B1-B2 transition and high-pressure melting temperature with the large amount of
theoretical and experimental work previously reported on. 
%--------------------------------------------------------------------

\section{Ab Initio simulations results}
\subsection{Computational details}
We carried out the {\sl ab initio} molecular dynamics simulations  using the ABINIT \cite{2009CoPhC.180.2582G} electronic
structure package. This consists in treating the electrons quantum mechanically using finite temperature density functional theory (DFT) while propagating
the ions classically on the resulting Born Oppenheimer surface by solving the Newton equations. We used the Local Density Approximation (LDA) with the Ceperley Alder
parameterization of the exchange correlation functional\cite{1980PhRvL..45..566C} that provides an equilibrium volume of 73.33$\AA^3$, closer to the experimental
measurement, 74.7 $\AA^3$, than in the generalized gradient approximation (GGA) for MgO\cite{2008JGRB..113.6204W}. While this somewhat arbitrary choice does not
guaranty a better behavior of the LDA formulation against the GGA one at high pressures, we point that sample calculations performed in the tens of Mbars range
indicate that the difference between the two tends to become less significant in the extreme pressure range investigated here.

We used two sets of projected augmented wave (PAW) pseudo-potentials to cover the entire density range considered here, from 6 to 22.5 g/cm$^3$.
For densities up to $9$g/cm$^3$, we used two projected augmented wave (PAW) pseudo-potentials generated by Jollet {\sl et al.}\cite{2014CoPhC.185.1246J}. These
pseudo-potentials are designed to reproduce accurately the all-electrons LAPW results obtained for the individual atomic species. This warranties that not
spurious effects due to the pseudization procedure are at play. For the two atomic species considered here, these pseudo-potentials consist in a PAW cutoff radius
of respectively 1.7a$_B$ and 1.2a$_B$. For both pseudo-potentials, the 1$s^2$ state only is kept as a core state while the remaining electrons are treated as valence
orbitals. To reach densities above 9g/cm$^3$ while insuring that the overlap of the PAW spheres remains marginal, we use the ATOMPAW \cite{2001CoPhC.135..329H} package
to generate pseudo-potentials with significantly smaller cutoff radius. For this second set of pseudo-potentials, the cutoff radius were fixed at r$_{paw}=1.1$a$_B$
and r$_{paw}=1.0$a$_B$ for, respectively, the magnesium and oxygen atomic species. The accuracy of the two set of pseudo-potentials produced was inferred by
comparing the T=0K EOS (cold curve) obtained for the B1 phase with the all electron LAPW results\cite{2008JGRB..113.6204W}. The static calculations were performed using the B1 unit
cell with a $8^3$ {\bf k}-points grid and a plane wave cutoff of 28Ha and 36Ha for, respectively, the soft and harder sets of pseudo-potentials.   

The molecular dynamics runs were performed using the finite temperature formulation of DFT as laid out by Mermin\cite{Martins} while the equations of motion
for the ions were integrated using the iso-kinetics ensemble. For each simulation runs, this consists in keeping the number of particles as well as the volume of
the simulation cell fixed while rescaling the atom velocities at each time step to keep the temperature constant. While it is well documented that this ensemble
is not formally corresponding to the canonical ensemble, it is used here out of convenience in a situation where the properties calculated such as the EOS or the
melting temperature are not sensitive to the use of a more refined thermostat \cite{tidsley}. We typically used a time step of 1fs at all thermodynamics conditions.
For the EOS calculations, we performed the simulations at the $\Gamma$-point and using 128 atoms in the simulation cell. We insured that the results obtained for the
pressure, internal energy, and ionic structure are converged to less than 1\% for both the B1 and B2 phases by doubling the number
of atoms and considering 256 atoms at a few sample P-T conditions. 
\begin{figure}
  \includegraphics[scale=0.4]{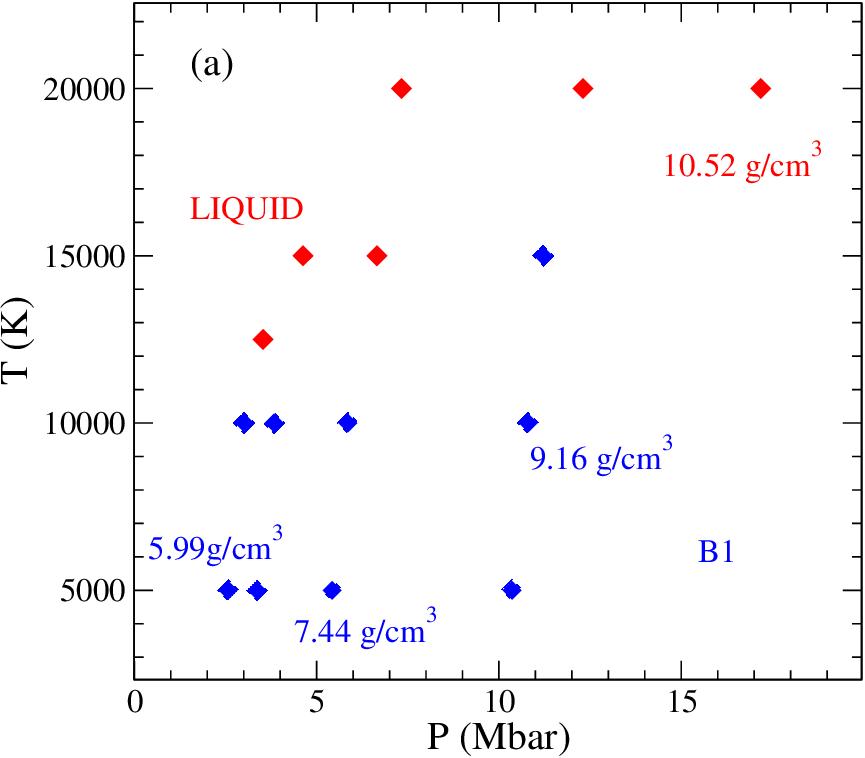}
   \includegraphics[scale=0.4]{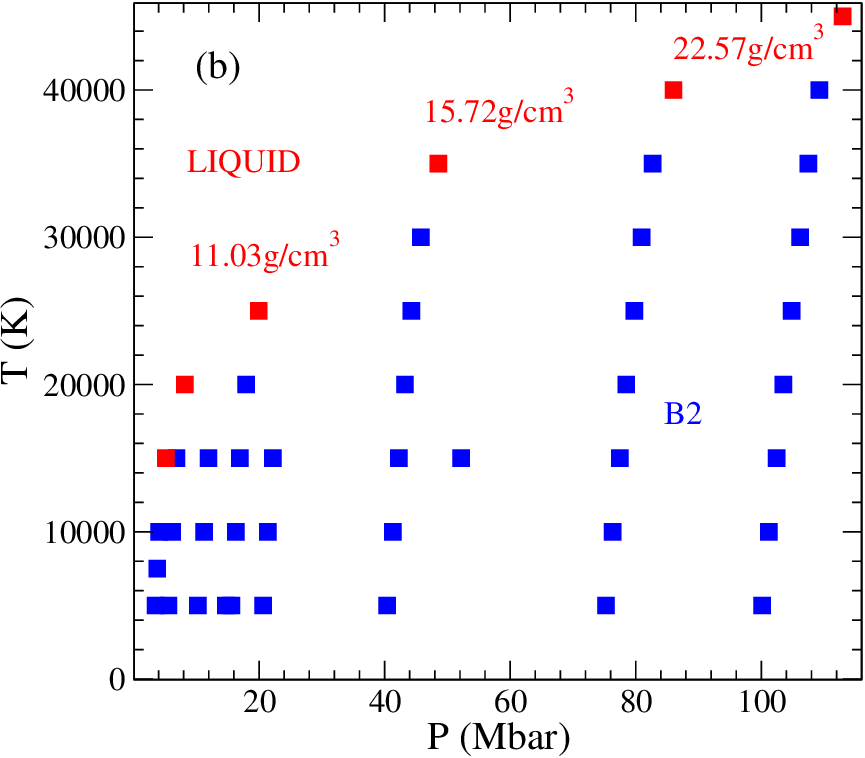}
   \caption{(a) EOS points obtained with simulations initially started in the B1 phase. (b) EOS points obtained with simulations initially started in the B2 phase.}
   \label{fig1}
\end{figure}
\subsection{Equation of States}

Fig. \ref{fig1}(a) and (b) show the EOS points obtained for simulations initially started in, respectively, the B1 and B2 phases. For the B1 phase, we calculate
EOS points using 128 atoms along six different iso-chores, 5.99g/cm$^3$, 6.45g/cm$^3$, 7.44g/cm$^3$, 9.16g/cm$^3$, 10.52g/cm$^3$, and 11.54g/cm$^3$
and for temperatures ranging from 5,000K to 20,000K. For the first four isotherms, we find results consistent with previous works and the
latest work of Cebulla {\sl et al.} who explored extensively this portion of the phase diagram below 10Mbar. We note that inspection of the correlation
functions indicates that the B1 phase is unstable on the two highest iso-chores.

Fig.\ref{fig1}(b) shows the EOS points obtained when starting the simulation in the B2 structure. To reach pressures relevant to the interior of giant planets, we
performed simulations along eight different iso-chores 6.76g/cm$^3$, 7.79g/cm$^3$, 9.64g/cm$^3$, 11.03g/cm$^3$, 12.20g/cm$^3$, 15.72g/cm$^3$, 20.10g/cm$^3$,
and 22.57g/cm$^3$ with temperatures varying from 5,000K to 45,000K. In contrast to the B1 structure that rapidly becomes unstable past the B1-B2 transition,
inspection of the correlation functions indicates that the B2 structure remains stable up to 120Mbar. While this does not formally warranty that the B2
structure is the lowest energy phase up to these extreme densities, the stability found combined with the high coordination number of Mg by oxygen in the B2 phase
is an indication that no other phase is expected beyond the B1-B2 transition.

In Fig.\ref{fig1}(a)-(b), we also display the state of the system at each pressure-temperature point by using the average mean square displacement. The average mean
square displacement is calculated using the two atomic species. We point out that close inspection of the mean square displacement for the individual
atomic species close to melting suggests that both participate in melting with no indication of partial melting on sublatticeis. Inspection of Fig.\ref{fig1}(a) and (b)
shows that the melting temperature rises rapidly to reach 40,000K for pressures beyond 100Mbar. It also shows that each phase exhibits a distinct pressure dependence
of its melting temperature. It is well documented\cite{2013PhRvB..87i4102B} that this basic approach leads to an overestimation of the melting temperature due to the
limited number of atoms that can be used in {\sl ab initio} simulations. We will turn below to the numerically intensive two-phase simulation method to reach a
more precise calculation of the melting temperature up to the conditions encountered in giant planet interiors and use this first estimation of melting as a boundary
to develop a simple parameterization of the {\sl ab initio} results.

For a convenient use of our {\sl ab initio} results in planetary modeling, we further adjusted a parameterization of the Helmotz free energy, $F(V,T)$ to the
{\sl ab initio} results. As developed further in Bouchet {\sl et al.}\cite{2013PhRvB..87i4102B}, this consists in expressing the Helmoltz free energy,$F(V,T)$, of the solid in
the adiabatic approximation as 
\begin{equation}
F=E_{cold}(V)+F_{harm}(V,T)+F_{anh}(V,T)+F_{el}(V,T)
\end{equation}
where $E_{cold}$ is the cold curve and represents the potential part of the free energy at zero temperature. $F_{harm}(V,T)$ is the harmonic phonon contribution that
can be obtained using linear response theory and the quasi-harmonic approximation for the ion motions, $F_{anh}(V,T)$  represents the correction due to anharmonics
effects and $F_{el}(V,T)$ represents the electronic contribution.

For the cold curve contribution, we use the Holzapfel form\cite{2001JPCRD..30..515H} as it provides the correct Thomas Fermi limit at infinite 
compression\cite{2010PhRvB..81e4110B}. As such, it is formally more appropriate than other parameterizations such as the Vinet or Birch-Murnaghan functional forms
to cover the compression range considered here. Within this parameterization, the pressure is given as:
\begin{equation}
P(V)_{cold}=3K_0X^{5}(1-X)exp\left[c_0(1-X)\right]\left[1+c_2X(1-X)\right]  
\end{equation}
with $X=\left(V/V_0\right)$, $c_0=-ln(3K_0/P_{FG0})$, $P_{FG0}=1003.6(Z/V_0)^{(5/3)}$, and $c_2=3/2(K'-3)-c_0$. $V_0$ is the equilibrium molar volume in cm$^3$/mole
while $P_{FGO}$ is given in GPa.

The thermal contribution is parameterized using the Einstein model where 
\begin{equation}
P_{harm}=\gamma E_{harm}/V
\end{equation}
with $ E_{harm}=3nR\left[\Theta/2+\frac{\Theta}{\exp(\Theta/T)-1}\right]$ and $R$ is the gas constant and $n$ the number of atoms, equal to two in the present case.
The Gruneisen parameter is given by
\begin{equation}
  \gamma=\gamma_{\infty}+(\gamma_0-\gamma_{\infty})(V/V_0)^\beta,
\end{equation}
where $\gamma_0$ and $\gamma_{\infty}$ represent its values at respectively ambient conditions and infinite compression following Al'tshuler \textit{et al.}
\cite{1987JAMTP..28..129A}. The Einstein temperature, $\Theta$ is expressed as
\begin{equation}
\Theta=\Theta_0(V/V_0)^{-\gamma_{\infty}}exp\left[\frac{\gamma_0-\gamma_{\infty}}{\beta}(1-(V/V_0)^\beta)\right],
\end{equation}
In these relations, $\Theta_0$, $\gamma_{\infty}$ $\gamma_0$ and $\beta$ are fitting parameters that formally do not carry any physical meaning.
Finally, as the anharmonic and electronic terms have the same temperature dependence\cite{2008PhRvB..78j4107B}, we used a single functional form given by 
\begin{equation}
P_{ae}=\frac{3R}{2V}ma_0x^mT^2,
\end{equation}  
to represent both contributions with $a_0$ and $m$ two fitting parameters. 

As for the case of iron\cite{2013PhRvB..87i4102B}, we fitted globally the {\sl ab initio} results using the complete functional form and without separating the
cold, $E_{cold}(V)$ contribution from the thermal one. This leads to the nine parameters given in Table\ref{tab1}.
\begin{table}[ht]
\begin{center}
  \begin{tabular}{|c|c|c|c|c|}
    \hline
 $V_{0}$(cm$^3$/mole) & $K_{0}$ (GPa) & $K'_{0}$(GPa)&$\Theta_0$ (K) &$\gamma_0$\\
\hline 
10.970  & 120.76      &   4.803 & 447.906  &  1.755\\  
\hline\hline
$\beta$&$\gamma_{\infty}$&$a_0$ (K$^{-1}$)&$m$&--\\
\hline
  -0.530  &   -7.579$\times10^{-2}$ &  -1.341$\times10^{-4}$ &  0.660& --\\
\hline
\end{tabular}
\end{center}
\caption{Set of parameters obtained by our fitting procedure of the B2 phase.}
\label{tab1}
\end{table}
This allows to reproduce the {\sl ab initio} pressure to an accuracy of less than 2\%. We give in Appendix 1 the {\sl ab initio} values used to performed the fit.
We note that we started the fit in the B2 phase slightly below the B1-B2 transition and for a density $\rho=6.51g/cm^3$. 
\subsection{High Pressure Melting}
We further used the computationally intensive two-phases method to evaluate the high-pressure melting temperature and correct for the over-heating
observed when considering direct bulk melting as used in the previous section. This simulation approach consists in using a super-cell initially
set up with both a liquid and a solid parts brought in contact and equilibrated in the iso-kinetic ensemble. In the results presented here, these initial
conditions were created by considering the simulation results obtained in the EOS calculations and using 128 atoms. Namely, along a given isochore, we set up a 256 atoms
simulation cell. We used the liquid and solid structures previously obtained at the same density and at the lowest temperature were melting was observed
and the highest temperature were the system remained solid. While the final results are not that sensitive to such a careful setup of the initial conditions, it remains
more efficient from a computational standpoint to start from partially equilibrated cells. As the direct bulk melting results provide an upper bound for the melting
temperature, the simulations were further started along each isochores with decreasing temperature in order to bracket the melting temperature.  All the
simulations used 256 atoms and were performed at the $\Gamma$-point. Previous studies\cite{2005PhRvL..94w5701A} show that this number of atoms is sufficient to obtain
accurate high-pressure melting properties. 
\begin{figure}
  \includegraphics[scale=0.4]{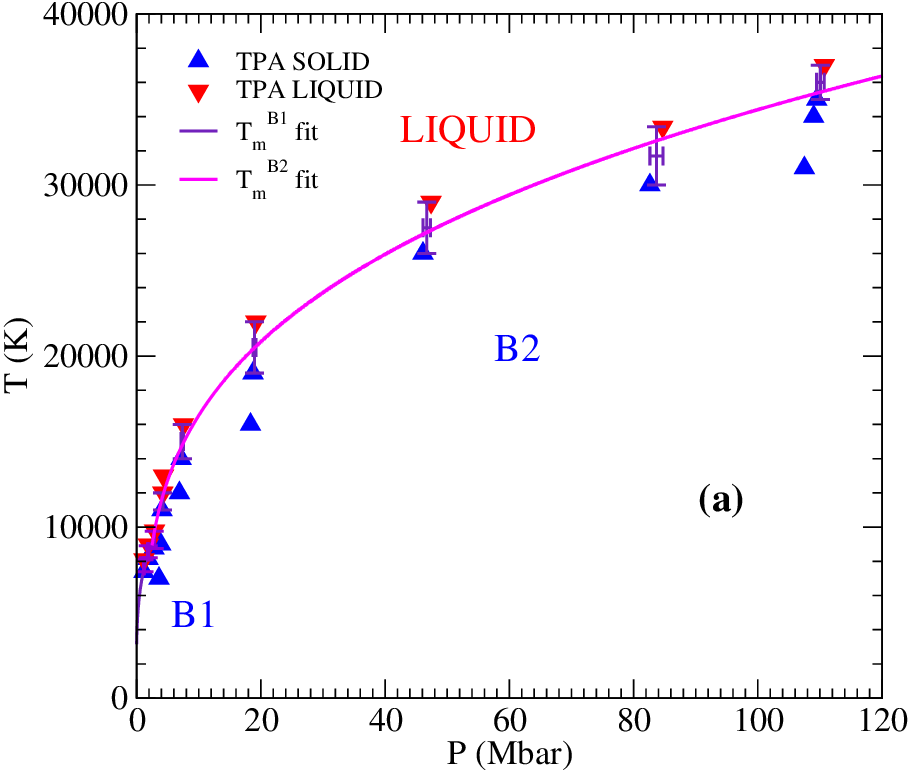}
   \includegraphics[scale=0.4]{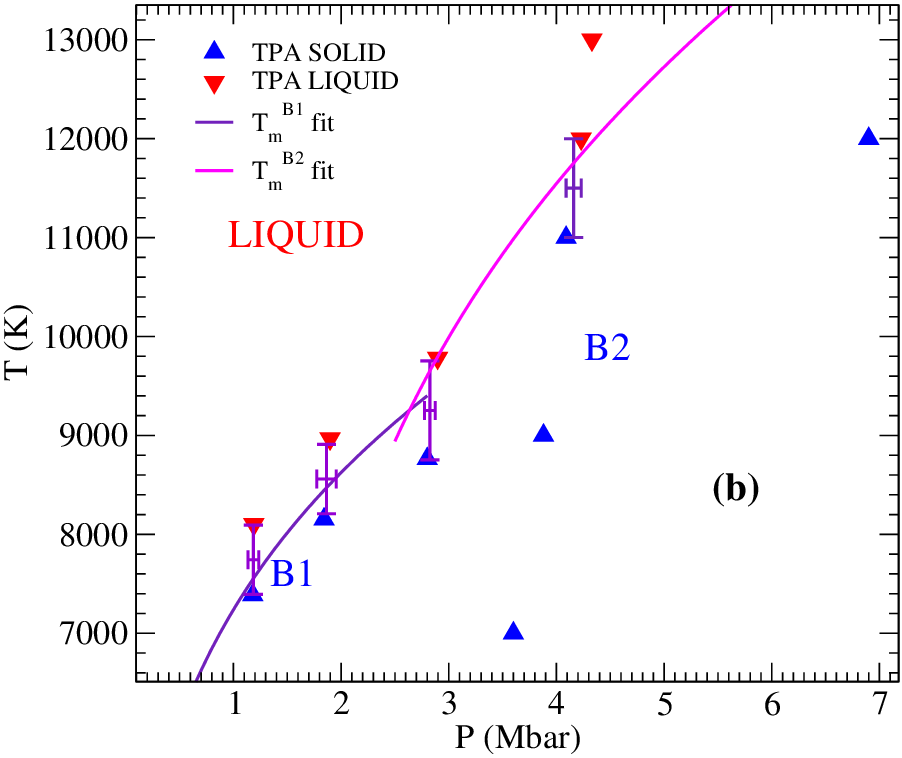}
  \caption{(a) High-pressure melting temperature obtained with simulations initially started using the B1 and B2 phases. (b) Zoom at pressures below 10Mbar. The color scheme indicates the final state after equilibration in the liquid (red) and in the solid (bleu).}
  \label{fig2}
\end{figure}

Fig.\ref{fig2}(a) and (b) show the results obtained, respectively, over the whole pressure range and at conditions below 7Mbar. As the melting temperature
is bracketed using this method, we take as melting temperature along an isochore as the mid-point between the lowest temperature where the simulation is
showing melting and the highest where it equilibrates to a solid. We further associate to this melting temperature error bars that
correspond to the average of these closest values. In effect, this corresponds to the uncertainty on the high-pressure melting temperatures
calculated with this method as there is no warranties that the melting temperature fall exactly at the mid-point. Fig.\ref{fig2}(a) shows a rapid increase of the
melting temperature in the B2 phase that reaches 34,000K for pressures above 100Mbar. We also see that the direct bulk melting method used above overestimates
melting significantly. At 100Mbar, we find that the melting temperature predicted by the two phase method is lower by 14\%. We further note that the melting
temperature in the B2 phase follows a simple Simon law with a slope of the melting temperature steadily decreasing as the pressure increases.
\begin{table}[ht]
\begin{center}
  \begin{tabular}{|c|c|c|c|c|}
    \hline
 Phase & $T_{ref}$(K)& $P_{ref}$(Mbar)&$a$(Mbar)&$c$\\
\hline 
B1  & 3152& 0&0.043     & 3.83436 \\  
\hline
B1 & 9255& 2.6363& 1.25668& 3.32258\\
\hline
  \end{tabular}
  \caption{Coefficients of the Simon-Glatzel form fitting the {\sl ab initio} results in the B1 and B2 phases.}
  \label{tab2}
\end{center}
\end{table}
Fig.\ref{fig2} shows a fit of the {\sl ab initio} results using the Simon-Glatzel law for each of the two phases. This empirical law relates the melting temperature
$T_m$ to the pressure using the relation 
\begin{equation}
  T_M=T_{ref}\left(\frac{P_M-P_{ref}}{a}+1\right)^{(1/c)},
\end{equation}
where $T_{ref}$ and $P_{ref}$ are the temperature and pressure of the triple point while $a$ and $c$ are two adjustable parameters. For the B1 phase, we completed the
{\sl ab initio} two-phases simulation results with the two lower density points of Alfe \cite{2005PhRvL..94w5701A} and the experimental melting temperature at normal
conditions to perform the fit. For the B2 phase, we iteratively adjusted the reference point, which corresponds in this case to the position of the triple point, to
match the melting temperature of the B1 phase and provide at the same time an accurate evaluation of the melting temperature throughout the entire pressure range.
Using this approach, we estimate the location of the triple point at $(P_{ref}=2.6363Mbar;T_{ref}=9,255K)$. Table\ref{tab2} shows the coefficients obtained for each
phase. We now turn to comparing our results with previous estimations up to 10Mbar.

\section{Comparison with previous work}
\begin{figure}
  \includegraphics[scale=0.4]{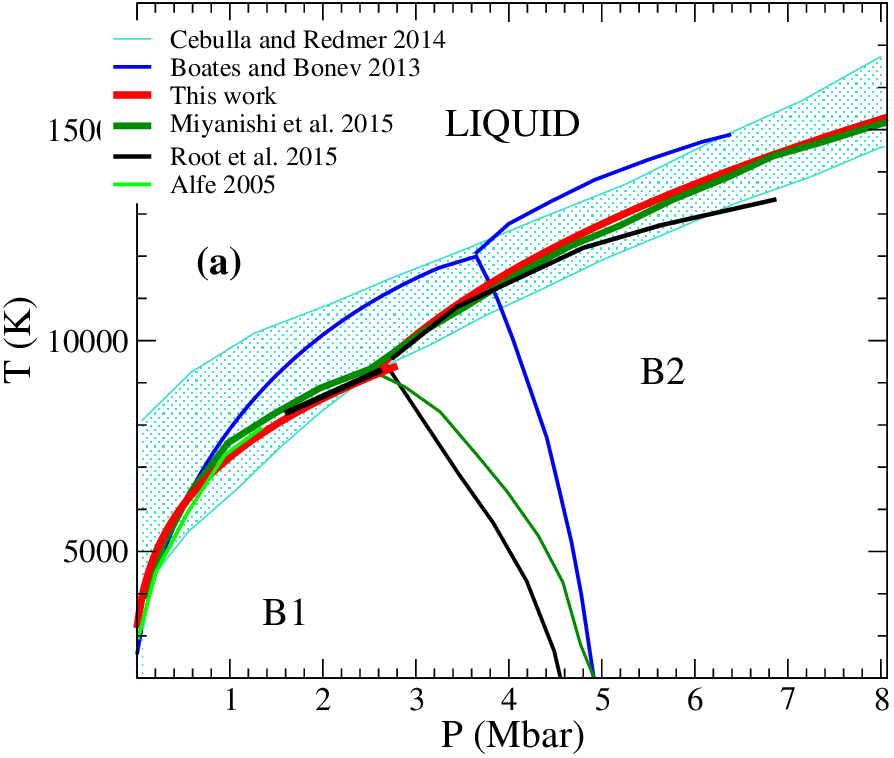}
   \includegraphics[scale=0.4]{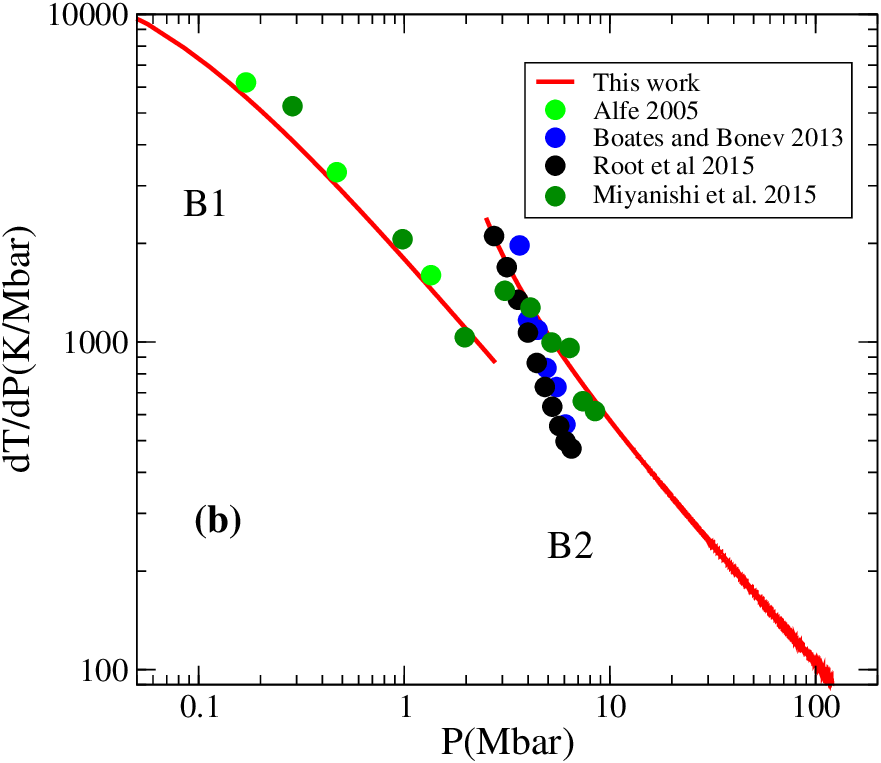}
  \caption{(a) Comparison between the high-pressure melting curve obtained in this work and previous results.  (b) Comparison between the slopes of the melting curve, $(d T/d P)$, obtained in this study with previous work.}
  \label{fig3}
\end{figure}
We compare in Fig.\ref{fig3} our estimation of the high-pressure melting curve for the B1 and B2 phases with the latest theoretical estimations. Following the
pioneer work of Alfe\cite{2005PhRvL..94w5701A}, who made the first calculation of the high-pressure melting temperature up to the pressure encountered at the Earth
core-mantle boundary using DFT based coexistence simulations, several predictions followed to extend this work beyond the B1-B2 transition. Fig.\ref{fig3}-(a) shows
that in the B1 phase, our prediction agrees nicely with the estimation of Root {\sl et al.}\cite{2015PhRvL.115s8501R} and Miayanishi
{\sl et al.}\cite{2015PhRvE..92b3103M} and contrasts with the one of Boates and Bonev\cite{2013PhRvL.110m5504B}. This latter prediction, while based on {\sl ab initio}
calculations relies on a unconventional approach to estimate the ionic entropy in the liquid. Our calculation in the B1 phase confirms that the prediction of Boates
and Bonev\cite{2013PhRvL.110m5504B} is unlikely beyond 1Mbar. A closer inspection shows slight differences of a few hundreds Kelvin between the three other
calculations but they remain well within the error bar of the method. This estimation of melting in the B1 phase validates our simulation parameters to predict the
high-pressure melting in the B2 phase.

Fig.\ref{fig3}-(a) shows that the comparison in the B2 phase is more contrasted. While our prediction of melting in the B2 phase agrees with the prediction of Root
{\sl et al.}\cite{2015PhRvL.115s8501R} and Myanishi {\sl et al.} up to 5 Mbar, we see a departure with the former at 7Mbar. The good agreement with the prediction
of Miyanishi {\sl et al.}\cite{2015PhRvE..92b3103M} up to 10Mbar suggests that this may come from the pseudo-potentials used as all the simulation parameters remain
otherwise the same. The differences between the various predictions can be better highlighted by considering the slope of the melting curve, $d T/d P$.

Fig.\ref{fig3}-(b) displays the slope of the melting curve as a function of pressure as obtained from the various predictions. We see, as expected, a satisfying agreement
between the various predictions for the B1 phase, with a slope continuously decreasing as the pressure increases. In the B2 phase, we see that all the predictions agree
reasonably above the B1-B2 transition occurring around 3Mbar and up to 6Mbar. In agreement with what we noticed previously for the melting curve above 6Mbar, we see
that the Root {\sl et al.}\cite{2015PhRvL.115s8501R} prediction for the slope of the melting curve departs from the prediction made in this work as well as with the
prediction of Miyanishi {\sl et al.}\cite{2015PhRvE..92b3103M} at higher pressures. This good agreement with the result of Miyanishi {\sl et al}\cite{2015PhRvE..92b3103M} up to the highest pressure they investigated gives us some confidence on the accuracy of our prediction at higher pressures and for the regime occurring within
giant planets.

Fig.\ref{fig3}-(b) indicates that the slope is steadily decreasing as pressure increases and reduces by close to an order of magnitude from the B1-B2 transition at 3Mbar
up to the regime relevant to giant planet interiors. For Jupiter, the largest planet in our solar system, this regime occurs between 40Mbar and 70Mbar. By considering
the Clausius-Clapeyron relation $dT/dP=v_m/s_m$, with $v_m$ and $s_m$ respectively the volume and entropy changes at melting. This continuous decrease of the slope
suggests that the volume change and probably also the entropy at melting continuously decreases as pressure increases. This suggests that MgO follows the standard
behavior of materials with a behavior in the B2 phase similar to what was pointed out for the B1 phase by Alfe\cite{2005PhRvL..94w5701A}. .    

We also stress that our estimation of the B2 high pressure melting below 5Mbar remains, within the error bar of the method, similar to the result of Root
{\sl et al.}\cite{2015PhRvL.115s8501R}. Furthermore, our crude estimation of the triple point, based on fitting the high pressure melting temperature in the B1 and B2
phases using a Simon-Glatzel law is also consistent with their result. As such, we confirm their analysis regarding the position of the B1-B2 transition and the
discrepancies with the experimental measurement of Coppari {\sl et al.}\cite{2013NatGe...6..926C}  who found the solid-solid B1-B2 transition occurring at 6Mbar for
temperatures of 5,000K. Similarly, our results fail to interpret the discontinuity observed in the experimental measurements of Mc Williams {\sl et al.} and Bolis
{\sl et al.} who both used decaying shocks to probe the equation of state of MgO at pressures around 5Mbar and temperatures close to 10,000K. The discontinuity
observed in the (P,T) space does not correspond to either the B1-B2 transition or the high pressure melting temperature as calculated by {\sl ab initio} simulations.
This analysis is thus not reproduced here. We point out however, that the analysis of the density of states and the calculation of the electrical conductivity using
the Kubo-Greenwood formulation\cite{2010HEDP....6...84M} in each of the three phases unambiguously indicates that MgO is a semi-conductor in the two solid phases and is metallic in the liquid phase\cite{riccardo-thesis}.

\section{Implication for planetary modeling}
\begin{figure}
  \includegraphics[scale=0.4]{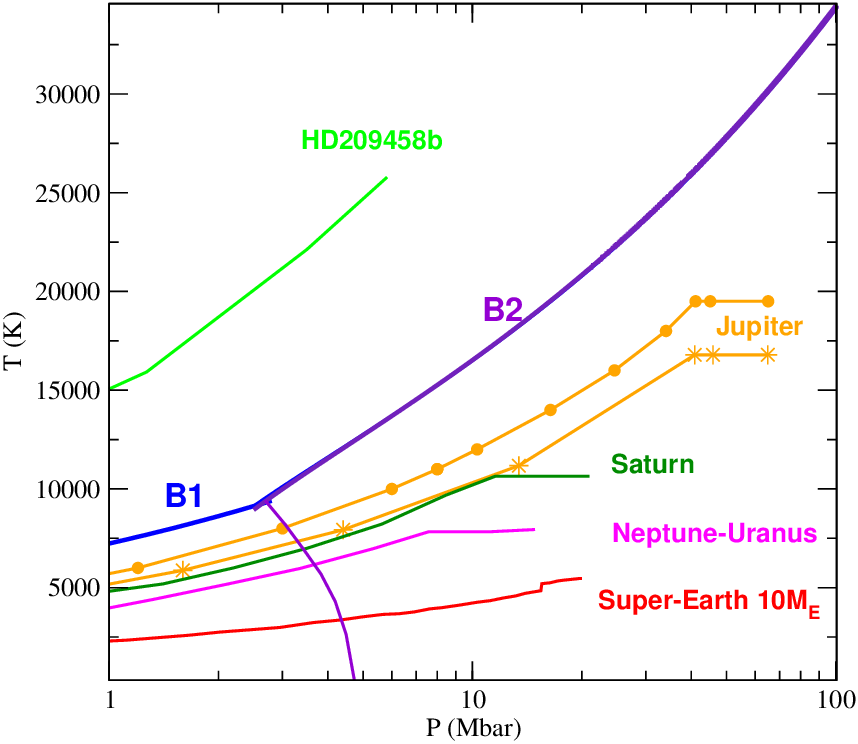}
  \caption{Comparison between the B2 high pressure melting temperature with planetary interior profiles.}
  \label{fig4}
\end{figure}

We compare in Fig.\ref{fig4} our prediction for the B2 high-pressure melting temperature with planetary interior profiles representative of the various objects
currently observed. These interior profiles are all obtained considering the planet as made of successive layers of varying composition in hydrostatic
equilibrium\cite{2014arXiv1405.3752G}. For Jupiter and Saturn, this mainly consists in an envelop made of hydrogen and helium with a core represented as a mixture
of water and silicates. For Neptune and Uranus, the paradigm consists in considering three layers, an H-He envelop, a large water mantle, and a central core made
of silicates. Super-Earths are approximated as a silicate mantle, that can contain a significant amount of MgO, with an iron inner core. Within this
picture, MgO is expected to be found either from the initial formation or as a dissociation product of MgSiO$_3$ or Mg$_2$SiO$_4$\cite{2006Sci...311..983U}. 

Fig.\ref{fig4} shows various interior profiles for these three types of planets. For Jupiter, we show the two latest estimations\cite{2016A&A...596A.114M} of Militzer
{\sl et al.} \cite{2013PhRvB..87a4202M} and Nettelmann {\sl et al.}\cite{2012ApJ...750...52N}. The plateau above 40Mbar represents the conditions encountered in the
inner core where an isothermal approximation is used when solving the hydrostatic equations. We see in Fig.\ref{fig4} that, if present, MgO would be found in a
solid and in the B2 phase. We indeed see that the conditions anticipated in the core by either models are well below the melting temperature predicted in the current
work. For pressures above 40Mbar, we predict melting temperatures steadily increasing from 25,000K. This is also the case for the other giant planet of our solar
system, Saturn, where the conditions at the core are predicted to be above 10Mbar and temperature around 11,000K\cite{2014arXiv1405.3752G}. The melting temperature
of the B2 phase is predicted to be close to 50\% higher and thus well above the core conditions.

As we turn to the two icy-giants, Neptune and Uranus, less extreme conditions prevail in the core\cite{2014arXiv1405.3752G}. The pressure is
estimated to be on the order of 7Mbar and higher and the temperature around 7,500K. The estimated temperature for the core is now a factor of two lower than the
predictions of the high-pressure melting temperature made in this work and corroborated by two other {\sl ab initio} calculations as described in the previous sections.
For the last type of planets, super-Earths, we show in Fig.\ref{fig4} a profile for a planet of 10M$_E$ (M$_E$ Earth mass)
\cite{2008PhST..130a4035S,2007ApJ...665.1413V,2011HEDP....7..141M}. The conditions encountered in the silicates mantle is estimated to be below 5,000K up to 15Mbar.
In Fig.\ref{fig4}, the discontinuity at 15Mbar represents the interface between the envelop (or mantle) and the core.

Fig.\ref{fig4} shows that the only system where MgO could be anticipated in a liquid state in the planet interior is a hot exoplanet such as HD209458b. Planets
of the solar system all show that MgO would be in a solid B2 state if present. This is also the case for super-Earths up to 10M$_E$. We also point out that the
result obtained in this work for the high-pressure temperature of the B2 phase also allows us to anticipate that this is likely the case for giant exoplanets several
times the size of Jupiter. Indeed, interior structure calculations for a planet three times the size of Jupiter and of comparable age indicate that the core spans
pressures from 300 to 400Mbar and temperatures around 40,000K. As we found by direct calculations that the melting temperature reached comparable value by 100Mbar
with a positive slope, we anticipate that MgO will be in the solid phase for these objects is the B2 phase is stable up to this point.     

\section{Summary and acknowledgements}
We calculated the EOS and high pressure melting curve of mgO in the B1 phase and B2 phase up to 120Mbar. Exhaustive comparison with previous work
up to 10Mbar shows satisfying agreement with previous estimations for both the EOS and the high-pressure melting temperature in both phases. We provide both quantities
in parametric form for a direct use in planetary modeling. Direct comparison with various estimations of the interior structures of the planets of the solar system
indicates that mgO is likely found in solid state and in the B2 phase within all these objects due the steep increase of the
melting temperature with increasing pressure.

Part of this work was supported by the ANR grant PLANETLAB 12-BS04-0015. Funding and support from Paris Sciences et Lettres (PSL) university through the project origins and conditions for the emergence of life is also acknowledged.

\bibliographystyle{aip} %
%\bibliography{bib-mgo} 

\end{document}